\documentstyle[aps,pra]{revtex}
\draft
\begin{document}

\twocolumn[\hsize\textwidth\columnwidth\hsize
\csname @twocolumnfalse\endcsname

\title{Consistent histories, the quantum Zeno effect, and time of 
arrival}
\author{I. L. Egusquiza$^1$ and J. G. Muga$^2$}
\address{$^1$Department of Theoretical Physics,
\\
$^2$Departament of Physical Chemistry,\\
The University of the Basque Country,\\
Apdo. 644, 48080 Bilbao, Spain}
\date{\today}

\maketitle

\begin{abstract}
We present a decomposition of the general quantum mechanical evolution 
operator, that corresponds to the path decomposition expansion, and 
interpret its constituents in terms of the quantum Zeno effect (QZE). 
This decomposition is applied to a finite dimensional example and to 
the case of a free particle in the real line, where the possibility of 
boundary conditions more general than those hitherto considered in 
the literature is shown. We reinterpret the assignment of consistent 
probabilities to different regions of spacetime in terms 
of the QZE. The comparison of the approach of consistent histories to 
the problem of time of arrival with the solution provided by the 
probability distribution of Kijowski shows the strength of the latter 
point of view.
\end{abstract}

\pacs{PACS: 03.65.-w
\hfill  EHU-FT/0002}
]

\section{Introduction}

The theoretical treatment of ``time observables'' is an important 
loose end of quantum mechanics.  An example of the problems 
encountered was formulated by Misra and Sudarshan in the form of a 
paradox \cite{Misra:1977by}.  They seeked the probability that an 
unstable particle decay at some time during an interval 
$\Delta=[0,t]$.  This has to be distinguished, and in general differs 
from, the standard quantum probability that the particle be found 
decayed at the instant $t$.  More generally, they also looked for the 
probability that a quantum system makes a transition from a preassigned 
subspace of states to the orthonormal subspace during a given period 
of time, further examples being the dissociation of a diatomic 
molecule, or arrival of a particle at a region of space.  Classically, 
we can ask whether a particle moving on a line is always to the same 
side of the point $x=0$, be it to the right or the left (but always to 
the right or always to the left), or if it crosses the $x=0$ point 
during $\Delta=[0,t]$.  What are the probabilities for the 
particle being always to the same side during $\Delta$ or for the 
particle crossing, according to quantum mechanics?

Since many experiments deal with such topics, and provide answers for 
them, we may 
expect that quantum mechanics should provide an unambiguous recipe to 
compute these probabilities.  However, the standard formalism, as 
found in all textbooks, tells us only how to evaluate expectation 
values and probabilities for a given instant of time, so these 
questions seem to pose the need for some extension of the standard 
rules.  Misra and Sudarsan attempted an apparently natural procedure: 
they modelled the continuous observation implied in these issues by 
a repetition of ideal first kind measurements in the limit of infinite 
frequency.  The consequence of such an interpretation of the 
continuous measurement, however, is that the system never abandons the 
original subspace (quantum Zeno effect).  Misra and Sudarshan 
considered the contradiction between the theoretical prediction and 
actual experiments detecting time distributions (of arrival, of decay, 
in general of occurrence of events) as paradoxical.  For these 
authors, such a mathematical result was physically inacceptable: it 
was merely an indication that the assumed procedure was not adequate 
to provide the probabilities they were looking for.  The completeness 
of the quantum theory was therefore pending until a {\it trustworthy 
algorithm} could be found.  In fact, most of the many publications on 
the Zeno effect have been devoted to the analysis or implementation of 
the repeated measurement scheme, overlooking the origin of the 
paradox, namely the need to find a trustworthy algorithm for time 
distributions.

Later on, the formulation of non-relativistic quantum mechanics in 
terms of sum-over-histories opened up the possibility that some 
questions, even though lying outside the realm of the standard rules 
of quantum mechanics, could be sensibly posed.  One such question, for 
example, is whether it is possible to define probabilities for 
alternative regions of spacetime from amplitudes built as sums over 
restricted classes of paths.  This was indeed first discussed by 
Feynman himself \cite{Feynman:1948aa}.  Hartle \cite{Hartle:1991qg} 
and Yamada and Takagi \cite{Yamada:1991ts} \cite{Yamada:1996} studied 
the possibility to define a probability for crossing or not crossing 
$x=0$ in an interval $\Delta$ for a free particle on a line.  Their 
conclusion was that it is not possible to define such probabilities, 
because the interference term between the possibilities (the 
``decoherence functional'' of the literature) generically does not 
vanish.  Yamada and Takagi, however, pointed out that for 
antisymmetric initial wavefunctions it was indeed possible to define 
the probabilities, with the result that there was no crossing of the 
point $x=0$ whatsoever.  Another exception pointed out by Halliwell was 
the particle coupled to a bath.  However, the resulting probabilities 
depend on the nature of the bath and the coupling, i.e., no ideal 
distribution emerges.

In this paper we shall show how the class of states that allow for a 
positive answer within the consistent histories 
framework can be considerably enlarged.  This is done by means of a 
generalization of the PDX (path decomposition expansion).  The idea of 
summing over classes of (Feynman) paths is of course much more general 
than just its application to the example mentioned, and leads to many 
different interesting aspects.  One of particular interest to us, 
because of its possible relationship to the question of times of 
tunneling \cite{Fertig:1990} or arrival \cite{Yamada:1996b}, is the 
path decomposition expansion (PDX), first formulated by Auerbach and 
Kivelson \cite{Auerbach:1985rp} to study tunneling problems with 
several spatial dimensions.  We find the rather striking fact that, 
although hard wall boundary conditions have been assumed in 
all derivations of the decomposition formulae for the propagators, 
which is the central result obtained so far from the PDX, other 
boundary conditions could be imposed on the restricted propagator 
without impairing the validity of the expression.  

We shall start with an operator derivation of the PDX which is a 
generalization of the ones proposed by Halliwell 
\cite{Halliwell:1995jh} and Muga and Leavens \cite{gonrick:1999}.  As 
a simple illustration we shall apply it to a two-state system.  We 
shall then see that there is a set of exclusive alternatives for which 
the formalism of consistent histories 
\cite{Griffiths:1984rx,Omnes:1988ek,Omnes:1992ag,Omnes:1994bk,Gell-Mann:1993kh} 
cannot generically give a set of probabilities.  This will be 
understood in terms of the quantum Zeno effect for the two state 
system (which is actually the one that pertains to the proposal of 
Cook \cite{Cook:1988} and has been realized experimentally 
\cite{Itano:1990}).  Even though the example corresponds to a finite 
dimensional Hilbert space, the derivation of the PDX holds formally 
for infinite dimensional Hilbert spaces as well.  However, topological 
considerations come into play, and we show the need to specify 
boundary conditions for the restricted propagator.  We then analyze 
the Yes/No question formulated by Hartle and Yamada and Takagi, and 
show that it is possible to define probabilities consistently for a 
much wider class of initial conditions than the antisymmetric one put 
forward by Yamada and Takagi.  We explain the result by analogy to the 
finite dimensional example given previously.

This extension however falls short of the broad generality that can be 
attributed to other conventional approaches, in particular to the 
definition of probabilities by means of positive operator valued 
measures: the time of arrival distribution of Kijowski is perfectly 
well defined for free particles on the line.  Our aim in the final 
discussion is to solve this apparent contradiction.

\section{Operator derivation of the PDX}

Halliwell \cite{Halliwell:1995jh} obtained an operator derivation of
the PDX which is closely related to the point of view of consistent or
decoherent histories
\cite{Griffiths:1984rx,Omnes:1988ek,Omnes:1992ag,Omnes:1994bk,Gell-Mann:1993kh}.
Let $P$ be a projector and $Q$ its complementary projector, $Q=1-P$.
Define $P(t)=\exp(i H t/\hbar) P \exp(-i H t/\hbar)=U^\dag(t) P U(t)$,
and similarly $Q(t)$.  It follows that if $H$ is self-adjoint,
$P(t)+Q(t)=1$ for every real $t$.  There exists a generalized
decomposition of unity, given by:
\begin{eqnarray}
	1 & = & P + \sum_{k=1}^n P(t_{k}) Q(t_{k-1}) Q(t_{k-2})\ldots
	Q(t_{1})Q\nonumber\\
	& & \quad +
	Q(t_{n}) Q(t_{n-1})\ldots Q(t_{1})Q\,,
	\label{ecdecomp}
\end{eqnarray}
for any set of real numbers $\{t_{1},t_{2},\ldots,t_{n-1},t_{n}\}$.
Assume that $t_{k}=k\delta t$, with $\delta t$ small.
Rewrite $P(t_{k})$ as
\begin{eqnarray}\label{desap}
	P(t_{k}) & = & P(t_{k-1}) + \delta t\,\dot P(t_{k-1})+ O(\delta t^2)
	\nonumber\\
	 & = & P(t_{k-1}) + \delta t\,U^\dag(t_{k-1})\dot P U(t_{k-1})+
O(\delta
	 t^2)\,,
\end{eqnarray}
where $\dot P$ is simply $\frac{i}{\hbar}[H,P]$. Multiply
(\ref{ecdecomp}) from the left with $U(t_{n})$, and use (\ref{desap}). We
obtain the following decomposition of the propagation operator:
\begin{eqnarray}
	U(t_{n}) & = & U(t_{n})P \nonumber\\
	& +& \sum_{k=1}^n \delta t\,U(t_{n}-t_{k-1}) \dot P U(t_{k-1})
	Q(t_{k-1})Q(t_{k-2})\ldots Q\nonumber\\
	&\, & + U(t_{n})Q(t_{n}) Q(t_{n-1})\ldots Q
	+O(\delta t^2) \,.
	\label{decompu}
\end{eqnarray}
Define the following ``restricted'' propagation operator
\begin{equation}
	U_{r}(t):=\lim_{n\to\infty,\delta t=t/n}U(n\delta t)Q(n\delta t)
	Q((n-1)\delta t)\ldots Q\,.
	\label{defur}
\end{equation}
Taking the limit $\delta t\to0$ in expression (\ref{decompu}) we
arrive at the generalized form of the PDX proposed by Halliwell
(see \cite{Halliwell:1995jh}, expression (2.19)):
\begin{equation}
	U(t)=U(t)P + \int_{0}^tds\,U(t-s)\dot P U_{r}(s) + U_{r}(t)\,.
	\label{decompucont}
\end{equation}
Notice that it can be further generalized without complication to time
dependent hamiltonians.

\subsection{Two state example}

Consider the two-state hamiltonian
$H=\hbar\omega\pmatrix{0&1\cr1&0\cr}$. Let $P=\pmatrix{1&0\cr0&0\cr}$,
and $Q=1-P$. The unitary evolution matrix is easily computed to be
\begin{equation}
	U(t) = \pmatrix{\cos(\omega t)& -i \sin(\omega t)\cr -i \sin(\omega
	t)& \cos(\omega t)\cr}\,.
	\label{unievol}
\end{equation}
It follows that $U_{r}(t)=Q$. Since $U(t)P=\pmatrix{\cos(\omega t)& 0\cr
-i \sin(\omega t)&0\cr}$ and $\dot P=\omega\pmatrix{0&-i\cr i&0\cr}$,
we see that each of the terms in (\ref{decompucont}) is different from
zero: the operator form of the PDX is not a trivial identity.

To interpret each of these terms, observe that $U_{r}(t)$, the
restricted propagation operator, corresponds to the continuous limit
of a series of preparations of the system in the subspace of states
invariant under $Q$. These preparations are equally spaced in time,
and are von Neumann collapses onto the eigenspace of $Q$. It is to be
expected, therefore, that this term is the propagator for a system
that is continuously observed in the eigenspace of $Q$, and this is,
in fact, the purport of the analysis of Misra and Sudarshan
\cite{Misra:1977by} of the quantum Zeno effect.

\subsection{Quantum Zeno effect}

If the initial state were in the eigenspace of $Q$, the term $U(t)P$
would not contribute to the later evolution of the system.  We
understand therefore that the convolution integral is the term
required to retain a probability that the initial quantum state in the
eigenspace of $Q$ does indeed jump at some point in time to the
eigenspace of $P$.  It is immediate to observe that the sum of the
convolution integral and the restricted propagator preserves the norm
of a state initially in the eigenspace of $Q$.  The quantum Zeno
effect can be understood in this case, therefore, as the decomposition of the
unitary evolution in the whole Hilbert space of an eigenstate of $Q$
in two terms: on the one hand the restricted propagator, which is
unitary in the eigenspace of $Q$, but non-unitary over the whole
Hilbert space, and on the other hand, the crossing term, necessary
to recover unitarity over the Hilbert space, and which accounts for
transitions out of the initial eigenspace.

Let us now pose the following questions: given a time interval $t$, and
a particle initially prepared with spin down (i.e., in the state
$|\downarrow\rangle=\pmatrix{0\cr1\cr}$), what is the probability that
it has always stayed with spin down in the interval?  What is the
probability that it has switched spin at some instant?  We can answer
the first one by looking at the restricted propagator $U_{r}(t)$:
the probability amplitude that it has always stayed with spin down is
$\langle\downarrow|U_{r}(t)|\downarrow\rangle=1$.  However, notice
that $\langle\uparrow|\int_{0}^tds\,U(t-s)\dot P
U_{r}(s)|\downarrow\rangle = -i \sin(\omega t)$ and
$\langle\downarrow|\int_{0}^tds\,U(t-s)\dot P
U_{r}(s)|\downarrow\rangle =\cos(\omega t) -1$.  It follows that we
cannot assign probabilities consistently to the exclusive events (i)
staying with spin down during the whole interval $t$; (ii) having
flipped spin at some instant of the interval.  The histories into
which we have decomposed the \emph{unitary} evolution of the particle
with initial spin down are not consistent histories!

In terms of operators, the operator associated to continuous 
measurement of being in the eigenspace of $Q$ and the operator 
associated with, at some point, jumping to the eigenspace of $P$ do 
not commute and give rise to a crossing term: they cannot be 
measured simultaneously.

More explicitly, the history operator associated with the particle 
always being in the eigenspace of $Q$ is 
$C_{1}=\lim_{n\to\infty,\delta t=t/n}Q(n\delta t)Q((n-1)\delta 
t)\ldots Q$, i.e.  a product of succeeding projectors.  The 
complementary operator is $C_{2}=1-C_{1}$.  The decoherence functional 
is $d(i,j)={\rm{Tr}}\left(C_{i}\rho C_{j}^\dagger\right)$, and the 
inconsistency of probability assignments is reflected in the fact 
that, in the case portrayed above, 
${\rm{Re}}\left(d(1,2)\right)\neq0$.  Notice that the history operator 
$C_{1}$ is related to the restricted propagator defined above through 
the following expression: $C_{1}=U^\dagger(t) U_{r}(t)$.

It is relevant at this point to mention the ``spectral decomposition'' 
approach of Pascazio and Namiki \cite{Pascazio:1994}, 
similar to the idea of the generalized PDX presented above. 
Additionally, notice that the models in the literature that attempt to 
obtain the quantum Zeno effect as a consequence of decoherence are in 
fact cancelling out the crossing term. In other words, if the 
pointer basis for a decoherence process is adequately aligned with 
the eigenspaces of $P$ and $Q$, the quantum Zeno effect will be 
immediately obtained as a consequence of decoherence. 

Insofar as the quantum Zeno effect is a paradox (see \cite{Home:1997}
for a general discussion), it is a paradox in that what seem to be
exclusive and consistent events for assignments of probability in
classical mechanics cannot be assigned quantum mechanical
probabilities in a consistent manner.  It should be stressed however
that this is no logical internal contradiction of quantum mechanics.
Rather, this simply reflects the fact that statements about quantum
events have to be much more precisely enunciated, and that classical
language and presuppositions do not always translate readily into the
quantum world.

\section{Histories on the real line}

The derivation of (\ref{decompucont}) presented above is formal, with
no attention being paid to topological issues.
In order to highlight the difficulties, consider the case of a free
particle of mass $m$ that moves on a line.  By simple
integration by parts one can realize that $PHQ$ need not be zero, since
\begin{equation}
\left(PHQ\psi\right)(x)=\frac{-\hbar^2}{2m}(1-\theta(x))\partial_{x}^2
\left(\theta(x)\psi(x)\right)\,.
\end{equation}
It therefore behooves us to analyze the meaning of $U_{r}(t)$. It is
obtained as a time ordered limit of products of $QHQ$ terms. The
operator $QHQ$, however, is not
self-adjoint: it admits a continuous one parameter family of self-adjoint
extensions. Therefore, unless a particular self-adjoint extension is
chosen, $U_{r}(t)$ will not be unitary in the eigenspace of the
projector $Q$. Imagine now that a particular extension has been chosen.
The meaning of $PHQ$ is subsirvient to the extension chosen, since
what we actually require is $PHQ + QHQ= HQ$. If the meaning of $QHQ$
is modified, so should the meaning of $PHQ$ be modified.

This observation can be strengthened by applying the theorem of Misra
and Sudarshan concerning the quantum Zeno effect \cite{Misra:1977by}
to this case of the free particle. The Hamiltonian of the free
particle is self-adjoint and semibounded (first assumption of the
theorem), and there
exists a time reversal operator, which commutes with the projectors
onto spatial regions (second assumption). Suppose now that the limit
defining $U_{r}(t)$ exists; actually assume that it exists in the
strong topology. It is clear in our case that if it does, its limit
when $t\to0$ is $Q$. It follows from Theorem 1 of ref. \cite{Misra:1977by}
that $U_{r}(t)$ then can be written as $Q\exp(-iBt/\hbar)Q$, with $B$
self-adjoint, and such that $QB=BQ=B$. The meaning of this result is
that the \emph{existence} of $U_{r}(t)$ implies the existence of a
self-adjoint operator to which it can be related, that can be
understood as a self-adjoint hamiltonian acting on the eigenspace of
$Q$. Therefore, the validity of the operator form of the PDX hinges
on choosing a specific self-adjoint extension of the original
hamiltonian when restricted to the $Q$-eigenspace, and considering the
unitary evolution in that subspace with this new hamiltonian.

Profiting from the simplicity of the example at hand, let us be more
specific.  The self-adjoint extensions of the free particle
Hamiltonian on the half-line are parameterised by a real parameter
$\beta$, and the domain of the extension $H_{\beta}$ is the set of
square integrable, absolutely continuous functions on the half line,
whose derivative is square integrable, and that fulfill the condition
$\psi(0)=\beta \psi'(0)$.

Thus the term $\dot P U_{r}(t)$ can be understood in terms of
integration by parts, as follows.  Define (formally) the propagator
$g(x,y,t)=\langle x|U(t)|y\rangle$ and the restricted propagator
$g_{r}^\beta(x,y,t)=\langle x|U_{r}^\beta(t)|y\rangle$, where
$U_{r}^\beta(t)=\exp(-i H_{\beta} t/\hbar)Q$.  The convolution integral
in (\ref{decompucont}) is then written as
\begin{eqnarray*}
	\langle x|\int_{0}^t&&ds\,U(t-s)\dot P U^\beta_{r}(s)|y\rangle =   \\
	 = \int_{0}^t&&ds\, \int_{-\infty}^{+\infty}d\xi\, g(x,\xi,t-s)
	 \theta(-\xi)
	 \left(\frac{-i\hbar}{2m}\right)\partial_{\xi}^2g_{r}^\beta(\xi,y,s)  \\
	 =&  & \left(\frac{-i\hbar}{2m}\right) \int_{0}^tds\,
	 g(x,\xi,t-s) \stackrel{\leftrightarrow}{\partial}_{\xi}
	 g_{r}^\beta(\xi,y,s)\Big|_{\xi=0}\,,
\end{eqnarray*}
where
$f(\xi)\stackrel{\leftrightarrow}{\partial}_{\xi}g(\xi)=
f(\xi)g'(\xi)-f'(\xi)g(\xi)$. It is important to stress that this
derivation is valid for all real $\beta$, not just for $\beta=0$,
which is the case analyzed in the literature.

As a matter of fact, Auerbach and Kivelson \cite{Auerbach:1985rp}
arrive at this symmetric form (with
$\stackrel{\leftrightarrow}{\partial}_{\xi}$ instead of
$\partial_{\xi}$) from the consideration that there is a change of
variable in the functional integral, trading $x_{\sigma}(s)$ for the
time $s$ after which the path is confined to one side of $x=0$, and
that the jacobian associated with this change of variables leads to
the symmetric operation $\stackrel{\leftrightarrow}{\partial}_{\xi}$.
However, they do not consider general boundary conditions of the form
stated here, because they do not seem to appear in their derivation of
the PDX in terms of a skeletonization of the path.  Other alternative
derivations \cite{Halliwell:1992nj,Fertig:1993} use Wick rotation, and
the diffusion process cannot see as physical alternatives all the
alternative boundary conditions that mantain unitarity for the
Schr\"odinger equation (in order to check this statement, see
\cite{Clark:1980ms} for the derivation of the restricted propagator in
the half-line through analytic continuation).  Hartle (see
\cite{Hartle:1991qg}, subsection 6.c and note 27) is rather
cautious in his analysis of Trotter's formula, which is basically what
underlies the definition of the restricted propagator, but is misled
by the uniqueness results available for the associated diffusion
equation.  Yamada \cite{Yamada:1996b} derives the PDX decomposition
out of a postulated integral equation, and imposes a particular choice
of boundary conditions, also missing out the alternatives highlighted
in the discussion above.

\subsection{Consistent probabilities}

Let us now ask the question first posed by Hartle \cite{Hartle:1991qg}
and, independently, Yamada and Takagi \cite{Yamada:1991ts}.  Is it
possible to assign consistently probabilities to the following
exclusive events: (i) that a free particle moving on the line stays
always to the same side of $x=0$ during a time interval $t$; (ii) that
it crosses $x=0$ once or more during the same time interval?  To make
the discussion easier, imagine first an initial wavefunction
restricted to the positive half-line.  Under the restricted evolution
$U_{r}^\beta(t)$, this wavefunction stays always in the positive
half-line with no loss of probability: $U_{r}^\beta(t)$ is unitary
when acting on $L^2({\bf R}^{+})$.  However, when we try to 
understand $U_{r}^\beta(t)$ as extended to an operator on
the whole real line, it is no longer unitary:
the convolution integral is required to guarantee the unitary
evolution of the initial one-sided state in the whole Hilbert space.
There is therefore a crossing term, and this prevents the
consistent assignment of probabilities to the exclusive events
mentioned.  As we see it, the requirement that a particle always be to
one side of the $x=0$ point is, in a way, imposed by constantly
monitoring that the particle is to one side, thus preventing the
classical exclusive events from being consistently exclusive also from
the quantum point of view.  In other words, we again run into the
quantum Zeno paradox.

Having said this, there \emph{is} an example of initial conditions, as
pointed out by Yamada and Takagi \cite{Yamada:1991ts}, for which the
probability assignments are consistent: the antisymmetric case.
Antisymmetric wavefunctions preserve this characteristic under
evolution with the free particle hamiltonian, or, in other words, the
parity operator commutes with the free particle hamiltonian.  This can
also be understood with regard to the restricted propagators as
follows: the evolution of an antisymmetric wavefunction under the
whole hamiltonian is identical to direct sum of the evolution in each
of the half-lines under the half-line free particle propagator with
hard wall boundary conditions.  There is no probability flow from one
half-line to the other under free-particle evolution if the initial
condition is antisymmetric.  This implies that in this case the
interference term is zero, and that the probability of always staying
to the same side during any time interval is unity: for any given
instant there is no probability of crossing $x=0$.

Given this point of view, it is immediate to generalize the example of
Yamada and Takagi to other instances: the meaning of the boundary
conditions that correspond to self-adjoint extensions of the free
particle hamiltonian when restricted to the half-line is that they
prevent probability flowing out of the half line. So for each real
$\beta$ we see that the wavefunctions that fulfill
$\psi(0)=\beta\psi'(0)$ have no transfer of probability from one half
line to the other. Alternatively, the evolution under the whole
hamiltonian of a wavefunction
obeying this condition is identical to the independent evolution of
the parts of the wavefunction in each of the half-lines under the
half-line free particle propagator with the corresponding boundary
conditions.
Thus we see that, for these initial wavefunctions, the assignment of
probability one to always staying to one side of the origin, and zero
probability to crossing the origin once or more during a time
interval, is indeed a consistent assignment of quantum probabilities.

\subsection{Arrival probabilities}

As seen above, only in some rather special circumstances can we make
consistent assignments of probability using a decomposition of
possible paths for the alternatives considered.  This does not mean,
though, that there is no consistent prescription within the realm of
standard quantum mechanics for the probability of having crossed a
given point, $x=0$, say, in a particular time interval.  Misra and
Sudarshan, in their seminal paper \cite{Misra:1977by}, already point
out that the existence of such a probability would imply the existence
of a generalized resolution of the identity (in their language; a
positive operator valued measure, or POVM, in modern parlance) for a
time of arrival operator. In fact, we now have at our disposal such a
POVM for the case of a free particle; the associated probability
density is, for a pure state $\psi$,
\begin{eqnarray}
\Pi_{K}(t,\psi) & = & \left|\int_{0}^\infty{\rm d}p\,\left(\frac{p}{2\pi
m\hbar}\right)^{1/2} e^{-ip^2 t/2m\hbar}\psi(p)\right|^2
+\nonumber\\ & &\quad
\left|\int_{-\infty}^0{\rm d}p\,\left(\frac{-p}{2\pi
m\hbar}\right)^{1/2} e^{-ip^2 t/2m\hbar}\psi(p)\right|^2\,,\nonumber
\end{eqnarray}
where we have used the momentum representation.  This is actually the
probability density proposed by Kijowski from an axiomatic point of
view \cite{Kijowski:1974}, which is related to the time of arrival
operator of Aharonov and Bohm \cite{Aharonov:1961} (see
\cite{Egusquiza:2000a} for details of the relationship between the two
objects).

Given this distribution, it is sensible to ask whether a similar 
construction could hold for the finite dimensional example given above. 
Unfortunately, the answer is negative. Imagine that indeed there 
exists a distribution of probability for the time of first shifting 
from $|\downarrow\rangle$ to $|\uparrow\rangle$. The existence of 
this distribution would imply the existence of a POVM (which in this 
finite dimensional example would have to be a projection valued 
measure, PVM), whose first operator moment, $T$ would be a 
self-adjoint operator (in this finite dimensional case, all symmetric 
operators are self-adjoint). Since this operator would have a ``time'' 
interpretation, it would have to be canonically conjugate to the 
hamiltonian, $[H,T]=i\hbar$. In the example at hand, $H$ is 
proportional to $\sigma_{1}$, and all operators, such as $T$, can be 
written as $\alpha +\vec\beta\cdot\vec\sigma$, where the $\sigma_{i}$ 
matrices are Pauli's matrices. There are no four 
numbers $(\alpha,\vec\beta)$ such that a canonically conjugate $T$ 
can be obtained. Therefore, there is no analogue of Kijowski's 
distribution for this finite dimensional example, and, in fact, there 
is no analogue of Kijowski's distribution for any finite dimensional 
example.

\section{Conclusions}

The operator derivation of the PDX formula we have presented here has
allowed us to identify the paradoxical aspects of the quantum Zeno
effect of Misra and Sudarshan as being due to incompatible assignments
of probability to inconsistent histories. We have explicitly separated
the crossing term that leads to this inconsistency.

Feeding the well-known results of Misra and Sudarshan back onto the
PDX formula, it also obtains that, in cases such as that of a free
particle moving on the line, there are several different PDX
expressions, each one corresponding to a particular partial isometry, i.e.,
to a particular self-adjoint extension of the restricted hamiltonian.
Furthermore, we have analyzed for which cases the PDX probability
assignments for the alternatives of having or not crossed a given
point are consistent, extending the result of Yamada and Takagi to all
instances of boundary conditions for which there is no probability
flow through that point. In spite of this extension, no time-of-arrival
probability could be assigned to the overwhelming majority of possible
states within the consistent histories approach.

We remark that there is a different, fully consistent prescription for 
the probability of having crossed a given point in a certain time 
interval, given by Kijowski's distribution in the free case.  Notice 
that Kijowski's distribution is obtained in the context of (almost) 
completely standard quantum mechanics, the only extension needed 
thereof being that POVMs are accepted to describe observables.  How is 
this distribution compatible with the negative results obtained within 
the framework of consistent histories?  The consistent histories 
approach is actually much more demanding, since it requires the 
absence of interferences between the space-time histories to attribute 
them a classical-like status as alternatives that actually occur with 
certain probabilities.  Instead, the distribution of Kijowski should be 
regarded, from the perspective of the standard interpretation, as a 
``potentiality'', a distribution that a properly designed apparatus 
could measure.  Therefore no association with non-interfering histories 
is claimed or required.  The apparatus would actually be the ``best'' 
one, in the sense of providing a covariant distribution with minimum 
variance.  Of course a less than perfect apparatus would provide 
convolutions or deformed versions of $\Pi_K$.  Werner has described 
the family of covariant distributions, each representing a 
potentiality associated with a different measurement device, for 
states with positive momentum components \cite{Werner:1988}.  From a 
more technical point of view, the difference can be associated with 
the fact that Kijowski's distribution at time $t$ is the expectation 
value for $\psi(t)$ of a certain operator, a quantum version of the 
positive flux minus the negative flux \cite{gonrick:1999}.  It is thus 
{\emph not} related to expectation values of strings of operators that 
depend on different instants of time.  In a slightly facetious way, we 
might say that standard, old-fashioned quantum mechanics has the upper 
hand on the consistent histories formalism for this particular case.  
While Kijowski's distribution is ``ideal'', in the sense of depending 
only on the state of the particle, there are other approaches in which 
additional degrees of freedom for the apparatus and or the environment 
are included, that provide operational time-of-arrival distributions 
\cite{gonrick:1999}.  Again, these results are found without demanding 
any non-interference condition.  Halliwell in particular 
\cite{Halliwell:1999} has compared the distribution derived from an 
irreversible detector model with the one associated with consistent 
histories in the presence of a bath coupled to the particle, and has 
showed how in the decoherent histories approach the coupling with the 
environment destroys far more interference that is really needed in 
order to define the arrival time with the irreversible detector.
 
For most cases of practical interest $\Pi_K$ is approximately equal to 
the current density $J$.  The challenge now is to perform experiments 
able to realize the ``potentiality'' of Kijowski's distribution in 
``quantum'' regimes where it differs significantly from the current 
density.  In general one may expect to obtain convolutions depending 
on the particular apparatus response \cite{Allcock:1969}, see 
\cite{gonrick:1999} for a more detailed discussion of the 
interpretation of $\Pi_K$.

One may wonder if Kijowski's distribution is the key to the 
``trustworthy algorithm'' seeked by Misra and Sudarsan for arbitrary 
problems where a time distribution for the the passage between 
complementary subspaces is required.  Indeed, the existence of 
Kijowski's distribution opens up the possibility that similar 
constructions might be feasible for other situations where the 
histories analysis has not been able to live up to its full promise.
However, we have proved that no analogue of Kijowski's distribution can 
be constructed in the case of finite dimensional Hilbert spaces. The 
question as to the existence of ``trustworthy'' analogues of 
Kijowski's distribution for infinite dimensional situations remains 
an open question, which we hope will be settled in the affirmative in 
the future (see \cite{Baute:2000} for an extension of Kijowski's 
distribution in the case of one dimensional motion with potentials).

\acknowledgements

We thank L.J. Garay, J.L. Ma\~nes, and M.A. Valle
for useful discussions, and acknowledge support
by  Ministerio de Educaci\'on
y Cultura (PB97-1482 and AEN99-0315), and The
University of the Basque Country (grant UPV 063.310-EB187/98)

\end{document}